\documentclass{article} %%% use \documentstyle for old LaTeX compilers
\usepackage[a4paper]{geometry}
\usepackage[english]{babel} %%% 'french', 'german', 'spanish', 'danish', etc.
\usepackage{amssymb}
\usepackage{amsmath}
\usepackage{txfonts}
\usepackage{mathdots}
\usepackage[classicReIm]{kpfonts}
\usepackage{amssymb}
\usepackage{graphicx}
 
\graphicspath{ {C:/Users/kjl33/Desktop/Langevin Paper Figures/} }

\begin{document}

%\selectlanguage{english} %%% remove comment delimiter ('%') and select language if required

\noindent 

\noindent 

\noindent 

\noindent 

\begin{center}
\noindent \noindent \textbf{\large DATA CLUSTERING BASED ON LANGEVIN ANNEALING WITH A SELF-CONSISTENT POTENTIAL}

\noindent \textbf{}

\noindent KYLE LAFATA

\noindent Department of Physics, Duke University

\noindent Department of Radiation Oncology, Duke University

\noindent Medical Physics Graduate Program, Duke University

\noindent kyle.lafata@duke.edu

\noindent \textbf{}

\noindent ZHENNAN ZHOU

\noindent Beijing International Center for Mathematical Research, Peking University

\noindent zhennan@bicmr.pku.edu.cn

\noindent \textbf{}

\noindent JIAN-GUO LIU

\noindent Department of Mathematics, Duke University

\noindent Department of Physics, Duke University

\noindent jliu@math.duke.edu

\noindent \textbf{}

\noindent FANG-FANG YIN

\noindent Department of Radiation Oncology, Duke University

\noindent Medical Physics Graduate Program, Duke University

\noindent fangfang.yin@duke.edu

\end{center}

\noindent \textbf{}

\noindent \textbf{\large Abstract}

\noindent \textbf{}

\noindent This paper introduces a novel data clustering algorithm based on Langevin dynamics, where the associated potential is constructed directly from the data. To introduce a self-consistent potential, we adopt the potential model from the established Quantum Clustering method. The first step is to use a radial basis function to construct a density distribution from the data. A potential function is then constructed such that this density distribution is the ground state solution to the time-independent Schr\"{o}dinger equation. The second step is to use this potential function with the Langevin dynamics at sub-critical temperature to avoid ergodicity. The Langevin equations take a classical Gibbs distribution as the invariant measure, where the peaks of the distribution coincide with minima of the potential surface. The time dynamics of individual data points lead to different metastable states, which are interpreted as cluster centers. Clustering is therefore achieved when subsets of the data aggregate –- as a result of the Langevin dynamics for a moderate period of time –- in the neighborhood of a particular potential minimum. While the data points are pushed towards potential minima by the potential gradient, Brownian motion allows them to effectively tunnel through local potential barriers and escape saddle points into locations of the potential surface otherwise forbidden. The algorithm’s feasibility is first established based on several illustrating examples and theoretical analyses, followed by a stricter evaluation using a standard benchmark dataset.

\section{Introduction}

\noindent Searching for meaningful structure within hyper-dimensional datasets is fundamental to modern data science. Cluster analysis -- i.e., the grouping of similar data objects together based on their intrinsic properties -- is a common approach to understanding otherwise non-trivial data. Although data clustering is a hallmark of many fields (e.g., machine learning, data mining, pattern recognition, etc. \cite{ARTICLE:3, ARTICLE:4, ARTICLE:5, ARTICLE:6, ARTICLE:7, ARTICLE:8, ARTICLE:9}), it is, in general, an ill-defined practice \cite{ARTICLE:10, ARTICLE:11} that may benefit from physics intuition. In this paper, we propose a novel approach to data clustering based on Langevin annealing. Data points are modeled as particles satisfying the second order Langevin equations with a self-consistent potential, where the Langevin dynamics are used to stochastically propagate them on a potential surface. The time dynamics of individual data points lead to different metastable states, which are interpreted as cluster centers. Clustering is achieved when subsets of the data –- as a result of the Langevin dynamics –- aggregate in the neighborhood of a particular potential minimum. To demonstrate the novelty of such a technique, we provide the following in this paper: (1) a complete mathematical derivation of the proposed technique, and (2) a numerical validation of the proposed technique, using a common benchmark dataset.

\noindent \textbf{}

\noindent Briefly, the motivation behind our approach to clustering is as follows. A given dataset is sampled from a canonical ensemble, whose distribution takes the form of the classical Gibbs measure,
\begin{equation}{{\rho }_A\left(\mathbf q\right)=Z^{-1}e}^{-\frac{V\left(\mathbf q\right)}{k_BT}}\end{equation}

\noindent where, $k_B$ is the Boltzmann constant, $T$ is the absolute temperature, $Z$ is the appropriate partition function, $\mathbf q\in {\mathbb{R}}^k$ is a position vector in $k$ dimensions\footnote{ The number of dimensions in the problem is inherently defined based on the number of data attributes within a particular dataset.}, and $V\left(\mathbf q\right)$ is a potential function that reflects the density of the data distribution. The Gibbs distribution, ${\rho }_A$, also serves as the $\mathbf q$ marginal distribution of the invariant measure of the following second order Langevin equation,

\noindent 
\begin{equation}m\frac{d^2\mathbf q}{dt^2}=-\mathrm{\nabla }V\left(\mathbf q\right)-\gamma \frac{d\mathbf q}{dt}+\mathrm{\Gamma }\eta \left(t\right)\end{equation}

\noindent where, $m$ is the particle's mass, $\gamma $ is a damping coefficient, $\eta (t)$ is a k-dimensional noise term, and $\mathrm{\Gamma }\mathrm{=}\sqrt{{2\gamma k}_BT}$ is an amplitude scaling coefficient \cite{ARTICLE:12}. According to the Fluctuation-Dissipation Theorem \cite{ARTICLE:13, ARTICLE:14, ARTICLE:15}, the damping term will \textit{dissipate} kinetic energy as it evolves in time, which is balanced with corresponding \textit{fluctuations} in the form of Brownian motion. Langevin dynamics are equivalently formulated as the following stochastic differential equations of motion, 
\begin{equation}d\mathbf q=\mathbf pdt,\end{equation}

\noindent and
\begin{equation}d\mathbf p=-\mathrm{\nabla }V\left(\mathbf q\right)dt-\gamma \mathbf pdt+\sqrt{2{\gamma k}_BT}dB\end{equation}

\noindent where, the noise term is specified as k-dimensional Brownian motion, $B(t)$, $\mathbf p\in {\mathbb{R}}^k$ is a time-dependent momentum vector, and we let $m\mathrm{=1}$ for simplicity. Formally speaking, while the forces from the potential surface push data points towards potential minima, corresponding Brownian fluctuations allow them to jump small potential barriers and escape saddle points into locations of the potential surface otherwise forbidden. The key observation is the following. Although each trajectory is ergodic with respect to the classical Gibbs distribution regardless of where it is initiated, when the potential barrier between two portions of the configuration space is big, it is highly likely that a trajectory initiated at one portion of the space will stay within the same portion for a moderate period of time. Therefore, we are motivated to propose the following clustering mechanism. When the potential function $V\left(\mathbf q\right)$ has several distinct minima, if we propagate the trajectories starting from the data points of a given data set, with a high probability, those trajectories will aggregate into several groups within a short time. Further, if the profile of the potential function intrinsically indicates the density distribution of the data points, the data points that end up in the neighborhood of the same minima of $V\left(\mathbf q\right)$ after $\mathrm{O(1)}$ time can be considered to be in the same cluster. 

\noindent \textbf{}

\noindent Clearly, the mechanics summarized via Equations (1)-(4) require a potential function, $V$. While $V$ can take on any well-defined functional form, we opt in this paper to use the Schr\"{o}dinger equation to model a self-consistent potential based on the method developed by Horn and Gottlieb \cite{ARTICLE:16}. As such, Euclidean data are mapped into a Hilbert space of square intergrable functions \cite{ARTICLE:17} by associating with a given data set a corresponding wave function, $\psi \left(\mathbf x\right)$. One can then interpret $\psi \left(\mathbf x\right)$ as the ground-state solution associated with the quantum mechanical Hamiltonian operator,
\begin{equation}\mathcal{H}\psi \left(\mathbf x\right)={\mathrm{-}\frac{{\varepsilon }^2}{2}{\nabla }^2}\psi \left(\mathbf x\right)+V\left(\mathbf x\right)\psi \left(\mathbf x\right),\end{equation}

\noindent and inversely solve for the potential function, $V\left(\mathbf x\right)$, whose minima correspond to cluster centers. Here, $\varepsilon $ is a rescaled, dimensionless analog of the reduced Planck constant, $\hslash \to \varepsilon $, such that $\varepsilon \ll 1$ defines the semi-classical regime: a transition regime between quantum mechanics and classical mechanics \cite{ARTICLE:18, ARTICLE:19}. Minima of $V\left(\mathbf x\right)$ intuitively correspond to amplified regions of high probability density within the original data set, leading to its high-resolution functional representation \cite{ARTICLE:16}. 

\noindent \textbf{} 

\noindent The \textit{dynamic} nature of the proposed methodology is consistent with recent research interests that study clustering as a dynamic problem \cite{ARTICLE:20, ARTICLE:21, ARTICLE:22, ARTICLE:24}. In general, dynamic frameworks provide the unique capability to investigate \textit{how} clusters are formed, and may potentially provide important meta-stable structural information about a particular dataset. In particular, Dynamic Quantum Clustering (DQC) extends quantum mechanics intuition to a dynamic framework \cite{ARTICLE:24}. By coupling data points with Gaussian wave-packets in a Hilbert space of wave functions, DQC interprets the data as solutions to the time-dependent Schr\"{o}dinger equation,

\noindent \textbf{} 

\noindent 
\begin{equation}i\varepsilon \frac{\partial }{\partial t}\left|\left.\mathrm{\Psi }\right\rangle \right.=\mathcal{H}\left|\left.\mathrm{\Psi }\right\rangle ,\right.\end{equation}

\noindent such that clustering is achieved by temporally propagating the system based on $\mathcal{H}$, and tracking the wave-packets on their oscillatory trajectory about potential minima \cite{ARTICLE:25}. 

\noindent \textbf{} 

\noindent The relative distances between the centroids of different states changes with time according to Ehrenfest's Theorem, thus representing a measure of similarity between data points. Wave-packets which oscillate around the same minima of $V\left(\mathbf x\right)$ are considered to be in the same cluster. In general, deriving weights from distance metrics such as this has been a hallmark of data clustering algorithms. As the convergence of many points to a common location is a clear and intuitive indication of clustered data, many approaches advocate constructing similarity functions. This is a way to characterize the distance between data points, and is an essential aspect of DQC, as well as methods of spectral clustering \cite{ARTICLE:20, ARTICLE:48, ARTICLE:49}, diffuse interface models \cite{ARTICLE:46, ARTICLE:47}, and graph-clustering based on diffusion geometry \cite{ARTICLE:20, ARTICLE:21, ARTICLE:44, ARTICLE:45}. 

\noindent \textbf{} 

\noindent Evolving the quantum system according to Equation (6) leads to closed-system physics driven by a completely conserved Hamiltonian, such that clustering is only a transient behavior. In our methodology, while we adopt the method used in \cite{ARTICLE:16} to generate $V\left(\mathbf x\right)$ explicitly from the data, we propagate the trajectories based on Langevin equations. Compared to \cite{ARTICLE:24}, this leads to a fundamentally different dynamic scheme influenced by the interaction between data points and a temperature-characterized environment. In fact, we derive (in \textit{Section 2} of this paper) a critical temperature, $T_0$, in the regime that Equation (1) is asymptotically matched with the quantum mechanics formalism for probability density,  

\noindent 
\begin{equation}{\rho }_B\left(\mathbf x\right)={\left|\psi \left(\mathbf x\right)\right|}^2,\end{equation}

\noindent 

\noindent such that,
\begin{equation}{Z^{-1}e}^{-\frac{V\left(\mathbf x\right)}{k_BT_0}}\approx {\left|\psi \left(\mathbf x\right)\right|}^2.\end{equation}

\noindent In practice, it is possible that at such a critical temperature, clustering of data is unlikely due to relatively large stochastic fluctuations. However, at sub-critical temperatures (that is, when ${T<T}_0$ is sufficiently small), the time-evolution is more strongly influenced by dissipative and systematic forces\footnote{ Except for particles near stationary points of the potential surface.}. Accordingly, the position of each data point tends to first move \textit{towards}, and subsequently move locally \textit{around}, nearby potential minima. As will be demonstrated in this paper, this thermal annealing process is the primary phenomenon which results in decoding meaningful structure about a given dataset. 

\noindent \textbf{}

\noindent Data clustering is therefore a direct consequence of Langevin dynamics at sub-critical temperatures for a moderate period of time (in contrast to ergodic sampling, which is a long time asymptotic behavior). In fact, in the regime that ${T\ll T}_0$, momentum decays to a small magnitude due to nominal stochastic influence (i.e., minimal Brownian motion where $\mathrm{\Gamma }\mathrm{\ll }\mathrm{1}$), resulting in a very high likelihood of data-point localization. Appropriately adjusting the sub-critical temperature in the thermostat provides a means to tailor the clustering phenomenon to that of a particular data set. This subtle yet important distinction offers a couple of key advantages. First, the Langevin dynamics framework is fundamentally different from time-evolution based on Ehrenfest's Theorem. The latter results in oscillatory trajectories about potential minima, which can often lead to practically difficult implementation. Sub-critical Langevin dynamics, however, lead to damped trajectories towards potential minima, resulting in the intuitive aggregation of data points. Second, Brownian motion allows data points to thermally jump local potential barriers and escape saddle points into locations of the potential surface otherwise forbidden. Such a tunneling phenomenon makes it possible for nearly degenerate local minima of $V\left(\mathbf x\right)$ to merge into a single cluster. Datasets can therefore be explored with a high resolution kernel, while still maintaining a reasonably narrow impulse response. 

\noindent \textbf{}

\noindent Given the proposed dynamic framework, there is a natural transition away from cluster analysis based solely on geometric intuition. By interpreting the potential landscape from a kinematic perspective -- rather than solely as a geometric entity -- it can be easily generalized to higher dimensional space \cite{ARTICLE:26}. Under such a generalization, the landscape of $V\left(\mathbf x\right)$ is characterized by the net systematic force felt by each data point at a given location contributing to the overall Langevin dynamics. In particular, there are three important scenarios on the surface of $V\left(\mathbf x\right)$ where such an interpretation may be used to evaluate the dynamic response of a test particle. These scenarios, which are detailed in \textit{Section 2} of this paper, include (1) evaluation of a test particle at a saddle point on the surface of $V\left(\mathbf x\right)$; (2) evaluation of a test particle far from all stationary points of $V\left(\mathbf x\right)$; and (3) evaluation of a test particle approaching minima of $V\left(\mathbf x\right)$.

\noindent \textbf{}

\noindent The remainder of this paper proceeds as follows. In \textit{Section 2}, we detail the theoretical framework of our proposed clustering algorithm, define the spatial and thermal conditions for its effective implementation, and derive a critical temperature to match the quantum and classical perspectives of a given dataset. In \textit{Section 3}, we briefly demonstrate the machinery used to numerically integrate the Langevin dynamics that drives the clustering process. Numerical results are presented in \textit{Section 4}. Using a common benchmark dataset, several computational experiments are designed to evaluate the methodology and characterize its key parameters. Concluding remarks and future directions are provided in \textit{Section 5}.

\section{Data Clustering with Langevin Annealing}

\noindent In this section, we aim to establish a general mathematical framework for data clustering based on Langevin dynamics. The Schr\"{o}dinger equation is first used to model a self-consistent potential function that reflects the probability density of a given dataset. Langevin dynamics are then formulated for such a potential, where clustering is expected to be a relatively low-temperature phenomenon. Finally, an analytical relationship is derived between the scale of the potential function and the temperature in the Langevin thermostat. The authors stress that, while quantum mechanics intuition is borrowed to establish self-consistency, the novelty of this paper is in the Langevin annealing. That is, the approach can be applied to any well-defined potential function. 

\subsection{Constructing a self-consistent potential}

\noindent Given a dataset, ${\left\{{\mathbf x}_i\right\}}^N_{i=1}\subset {\mathbb{R}}^k$, each ${\mathbf x}_i$ vector is first mapped from its original k-dimensional Euclidean space into a Hilbert space of square integrable functions, 

\noindent 
\begin{equation}{\psi }_i\left(\mathbf x\right)={Ce}^{-\frac{{\left|\mathbf x-{\mathbf x}_i\right|}^2}{2{\sigma }^2}}={Ce}^{-\frac{{\left|\mathbf x-{\mathbf x}_i\right|}^2}{2\varepsilon }}\end{equation}

\noindent where, $\sigma $ is the resolution of the chosen radial basis function kernel whose variance, $\varepsilon ={\sigma }^2$, is interpreted as the semi-classical parameter (a rescaled analog of the reduced Planck constant), and $C$ is the appropriate normalization constant. Similar to \cite{ARTICLE:16}, a wavefunction is then constructed as the superposition of these N wave-packets, 

\noindent 
\begin{equation}{\psi }_0\left(\mathbf x\right)=C_N\sum^N_{i=1}{e^{-\frac{{\left|\mathbf x-{\mathbf x}_i\right|}^2}{2\varepsilon }}}\end{equation}

\noindent where, $C_N$ is a normalization constant ensuring that, 

\noindent \textbf{} 
\begin{equation}\left\langle {\psi }_0\mathrel{\left|\vphantom{{\psi }_0 {\psi }_0}\right.\kern-\nulldelimiterspace}{\psi }_0\right\rangle ={\left\|{\psi }_0\right\|}^2_{L^2}=1.\end{equation} 
\noindent \textbf{}

\noindent To measure the initial overlap observed amongst the set of wave-packets at a given $\sqrt{\varepsilon }$-resolution, we introduce a set of weight functions, 
\begin{equation}w_i\left(\mathbf x\right)=\frac{e^{-\frac{{\left|\mathbf x-{\mathbf x}_i\right|}^2}{2\varepsilon }}}{\sum^N_{j=1}{e^{-\frac{{\left|\mathbf x-{\mathbf x}_j\right|}^2}{2\varepsilon }}}}\ ,\ \ \end{equation} 
such that, 

\noindent 
\begin{equation}\sum^N_{i=1}{w_i\left(\mathbf x\right)}=1.\end{equation} 

\noindent \textbf{}

\noindent Next, by requiring ${\psi }_0\left(\mathbf x\right)$ to be the ground state solution to the semi-classical Hamiltonian \cite{ARTICLE:16},

\noindent \textbf{}

\begin{equation}H{\psi }_0\left(\mathbf x\right)\equiv \left[-\frac{{\varepsilon }^2}{2}{\mathrm{\nabla }}^2+V\left(\mathbf x\right)\right]{\psi }_0\left(\mathbf x\right)=E_0{\psi }_0\left(\mathbf x\right),\end{equation}

\noindent the following potential function can be inversely calculated up to a constant,

\noindent \textbf{}

\begin{equation}V\left(\mathbf x\right)=\frac{\frac{{\varepsilon }^2}{2}{\mathrm{\nabla }}^2{\psi }_0\left(\mathbf x\right)}{{\psi }_0\left(\mathbf x\right)}+E_0=\frac{-k\varepsilon }{2}+\frac{\sum^N_{i=1}{{{\left|\mathbf x-{\mathbf x}_i\right|}^2e}^{-\frac{{\left|\mathbf x-{\mathbf x}_i\right|}^2}{2\varepsilon }}}}{2\sum^N_{i=1}{e^{-\frac{{\left|\mathbf x-{\mathbf x}_i\right|}^2}{2\varepsilon }}}}+E_0\end{equation} 

\noindent \textbf{}

\noindent where, $E_0$ is the ground state energy\footnote{ The ground state energy can be regarded as a bias term and does not practically affect the mechanics of clustering.}. The potential function, written in terms of Equation (12), becomes,
\begin{equation}V\left(\mathbf x\right)=\frac{-k\varepsilon }{2}+\sum^N_{i=1}{w_i\left(\mathbf x\right)\frac{{\left|\mathbf x-{\mathbf x}_i\right|}^2}{2}}{+E}_0.\end{equation}

\noindent Both ${\psi }_0\left(\mathbf x\right)$ and $V\left(\mathbf x\right)$ are intimately connected to the original Euclidean dataset. Maxima of ${\left|{\psi }_0\left(\mathbf x\right)\mathrm{\ }\right|}^2$ correspond to densely populated regions within ${\left\{{\mathbf x}_i\right\}}^N_{i=1}$, hence why this mapping is a fairly common approach in conventional kernel-density estimation algorithms \cite{ARTICLE:27}. Conversely, minima of $V\left(\mathbf x\right)$ correspond to locations where the wave-packets are subject to high degrees of local attraction, and are therefore interpreted as cluster centers \cite{ARTICLE:28}. Further, ${\psi }_0\left(\mathbf x\right)$ and $V\left(\mathbf x\right)$ are both characterized by resolution-dependent density variations throughout the data, parameterized by $\sqrt{\varepsilon }$. In particular, when $\varepsilon \ll 1$, each wave-packet is observed at a semi-classical resolution. This is a transition regime between quantum mechanics and classical mechanics, where the position and momentum of each wavepacket is $x_i$ and 0, respectively\footnote{ Momentum is zero because we have chosen, without loss of generality, not to include a phase term. If a non-zero phase term was used, the total wavefunction would no longer necessarily be real-valued. For simplicity in the current paper, we only consider the real-valued wavefunction.}.

\subsection{Langevin annealing and data clustering}

\noindent One essential feature of semi-classical wave-packets (which is fundamental to the proposed clustering algorithm) is as follows. Up to the Ehrenfest time, $t_E\approx -{\mathrm{log} \varepsilon \ }$, the wave-packets can be interpreted as coherent states, and the quantum dynamics can be approximated using Newtonian physics \cite{ARTICLE:29, ARTICLE:30}. That is, the time-evolution of the wave-packets can be approximated by evolving their parameters. Specifically, if we consider the semi-classical time-dependent Schr\"{o}dinger Equation,
\begin{equation}i\varepsilon \frac{\partial }{\partial t}{\mathrm{\Psi }}_i\left(\mathbf x,t\right)=\left[-\frac{{\varepsilon }^2}{2}{\mathrm{\nabla }}^2+V\left(\mathbf x\right)\right]{\mathrm{\Psi }}_i\left(\mathbf x,t\right),\end{equation} 

\noindent \textbf{}

\noindent whose initial condition is given by, 
\begin{equation}{{\mathrm{\Psi }}_i\left(\mathbf x,0\right)=\psi }_i\left(\mathbf x\right)={Ce}^{-\frac{{\left|\mathbf x-{\mathbf x}_i\right|}^2}{2\varepsilon }},\end{equation}

\noindent then, ${\mathrm{\Psi }}_i\left(\mathbf x,t\right)$ can be approximated by a parameterized wave-packet whose phase-space coordinates, $\left[\mathbf q\left(t\right),\mathbf p(t)\right]$, are determined by the classical Hamiltonian equations, 
\begin{equation}\dot{\mathbf q}=\mathbf p,\end{equation} 

\noindent and
\begin{equation}\dot{\mathbf p}=-{\mathrm{\nabla }}_{\mathbf q}V\left(\mathbf q\right),\end{equation}

\noindent with initial conditions, $\left[\mathbf q\left(0\right)={\mathbf x}_i,\ \ \mathbf p\left(0\right)=0\right] $\footnote{ To maintain consistency between the quantum mechanical and statistical mechanical perspectives, we have introduced the following notation: The points ${\hat{x}}_i$ are used when referring to the original, static dataset, and the points $\hat{q}$, $\hat{p}$ are used when referring to the time evolution of the phase space via Langevin dynamics. Accordingly, ${\hat{q}}_i\left(0\right)={\hat{x}}_i$ indicates that ${\hat{x}}_i$ and ${\hat{q}}_i$ are essentially equal counterparts of the quantum and classical distributions, respectively.}.  

\noindent 

\noindent \textbf{}

\noindent In this work, however, we treat the data points as classical points in equilibrium with a heat bath, characterized by a temperature, $\mathrm{T}$. Subsequently, we require that the trajectory of the particles obey the following Langevin dynamics, 

\noindent 
\begin{equation}d{\mathbf q}_i={\mathbf p}_idt,\end{equation}

\noindent and
\begin{equation}d{\mathbf p}_i=-{\mathrm{\nabla }}_qV\left({\mathbf q}_i\right)dt-{\gamma \mathbf p}_idt+\sqrt{2\gamma {\beta }^{-1}}dB,\end{equation} 

\noindent \textbf{}

\noindent with initial conditions, $\left[\mathbf q\left(0\right)={\mathbf x}_i,\ \ \mathbf p\left(0\right)=0\right]$. Here, $B\left(t\right)$ is k-dimensional Brownian motion, $\gamma $ is a damping coefficient, and $\beta ={\left(k_BT\right)}^{-1}$, where $T$ is the absolute temperature of the system, and $k_B$ is the Boltzmann constant. For simplicity, we let $\gamma =1$, and choose to work at the characteristic scale where $k_B=1$. In general, propagation of the position and momentum, based on Equations (21) and (22), respectively, lead to ergodic trajectories with respect to Equations (1) and (16). Further, it can be easily shown that the analytic form of the systematic force term (i.e., the gradient of Equation (16) is composed of both a linear component and a harmonic trap,

\noindent \textbf{}

\begin{equation}\mathrm{F}\mathrm{\equiv }\mathrm{-}\mathrm{\nabla }V\left(\mathbf x\right)=-\left\{\sum{\mathrm{\nabla }w_i\left(\mathbf x\right)\frac{{\left|\mathbf x-{\mathbf x}_i\right|}^2}{2}}+\sum{w_i\left(\mathbf x\right)}\left(\mathbf x-{\mathbf x}_i\right)\right\}.\end{equation} 

\noindent \textbf{}

\noindent When $\varepsilon \ll 1$, as we shall show in the next section, there exists a critical temperature, $T_0$, such that the quantum probability distribution is asymptotically convergent to a classical Gibbs measure. The interaction between stochastic and systematic forces preserve a state of equilibrium that guarantees a sampling of the desired canonical distribution for a sufficiently long time. Whereas, clustering, which is a short time behavior, may not be easily observed at critical temperature due to Brownian motion with a relatively large amplitude. 

\noindent \textbf{} 

\noindent However, when $T<T_0$ is small enough, the interaction between the potential function and the Brownian motion leads to a different equilibrium state, uniquely characterized by a sub-critical temperature, $T$. As such, it is highly likely that each data point follows the gradient of the potential towards the nearest minima of $V\left(\mathbf x\right)$ where it is trapped. Data points which end up trapped in the same minima of $V\left(\mathbf x\right)$ are considered to be in the same cluster. Interestingly, this implies that data clustering is fundamentally driven by Langevin dynamics at a sufficiently low temperature, $\mathrm{T}$. Further, while the systematic force acts globally on the system to push data points \textit{downhill}, the Brownian motion allows them to thermally jump local potential barriers into locations of the potential surface otherwise forbidden. Such a tunneling phenomenon makes it possible for the merging of nearly degenerate local minima of $V\left(\mathbf x\right)$. This enables the data to be explored with a high resolution kernel, while still maintaining a reasonably narrow impulse response (therefore reducing some of the noise that Equation (10) may introduce during the radial basis function operation).  

\noindent \textbf{} 

\noindent To track the time-dependent changes in data overlap that transpires during the dynamic process described above, we introduce a weight matrix, $w_{ij}$, whose initial condition,

\noindent \textbf{}
\begin{equation}w_{ij}\left(t=0\right)=w_i\left({\mathbf x}_j\right),\end{equation}

\noindent is determined via Equation (12). Here, we recall that the points ${\mathbf x}_j$ are time-independent by definition. Adhering to notation, the weight matrix as a function of time is then, 

\noindent \textbf{}
\begin{equation}w_{ij}\left(t\right)=w_i\left({\mathbf q}_j(t)\right),\end{equation}

\noindent as the points ${\mathbf q}_j$ are defined to be time-dependent. A high degree of dynamic overlap occurs for data points that aggregate together into similar clusters, i.e., $w_{ij}\to 1$. Conversely, data points that are separated into different clusters produce negligible dynamic overlap, i.e., $w_{ij}\to 0$. 

\noindent \textbf{}

\noindent While the magnitude of Brownian motion is invariant, the systematic net-force changes according to Equation (23). Evaluation of the net-force acting on a particle can be used to characterize the dynamics at various key locations on a potential landscape. As previously suggested in the introduction, such interpretation of the potential function's landscape as a kinematic quantity is somewhat of a paradigm shift away from conventional function optimization based solely on geometric intuition. In particular, the following illustrating examples demonstrate the theoretical response of a test particle\footnote{ Here, we refer to a test particle as being a 1D proxy for a data point within the illustrating example presented in Figure 1.} to three archetypical potential function interactions:

\noindent 

\begin{enumerate}
\item  A test particle evaluated far from potential stationary points: Here, the dynamics are dominated by steep potential gradients and dissipative effects, leading to damped Newtonian mechanics. Accordingly, the particle's trajectory roughly follows the gradient of the potential.

\item  A test particle evaluated at a saddle\textbf{ }point: Here, Brownian motion most prominently influences local dynamics, \textit{kicking} the particle out of the stationary point and allowing it to continue its trajectory.

\item  A test particle evaluated near functional minima: Here, $\mathrm{\nabla }V\left(\mathbf x\right)\approx 0$, and the momentum of the particle becomes small due to damping effects. Dissipation and fluctuation are comparable in magnitude, and they dominate the dynamics about the minima leading to dynamic particle confinement. 
\end{enumerate}

\noindent 

\noindent The simulated response of a test particle to each of these situations is demonstrated in \textbf{Figure 1}, using test-functions as a proxy for the potential landscape. Each simulation was executed from $t=0$ to $t=1$ with a time-step of $dt=0.01$. The red stars indicate the starting location of each test particle (with zero momentum), and the black lines demonstrate their time-trajectories according to Equations (21) and (22). 

\noindent \textbf{} 

\noindent A hyperbolic paraboloid (\textbf{Figure 1A}) was used to test the dynamic response of a particle with zero momentum starting at the function's saddle point. In general, effectively escaping saddle points is a fundamental research topic in optimization problems \cite{ARTICLE:31, ARTICLE:32} and manifold learning \cite{ARTICLE:33}. The magnitudes of the systematic and stochastic forces (\textbf{Figure 1A, bottom}) are shown as a function of the time-evolution. To demonstrate their relative effects as a function of time, force magnitudes have been normalized relative to 100\% simulation completion. It is confirmed that systematic and stochastic influence are comparable when the particle is near the stationary point, i.e., the systematic contribution is essentially trapped within the bandwidth of the stochastic noise. However, the systematic force is the dominating effect as the particle eventually locates the stronger potential gradient, as indicated in \textbf{Figure} 1A at simulation time, $t={t}^{'}$. Inverted Gaussian functions (displayed as 2D projections) were used to study the dynamic response of a test particle located near minima (\textbf{Figure 1B}) and far from stationary points (\textbf{Figure 1C}). Clearly, particle confinement at the minima is achieved in both cases as a function of time-evolution. 

\noindent 

\begin{center}
%\noindent \includegraphics*[width=5.52in, height=3.61in, keepaspectratio=false]{image1}
\includegraphics[width=.75\textwidth]{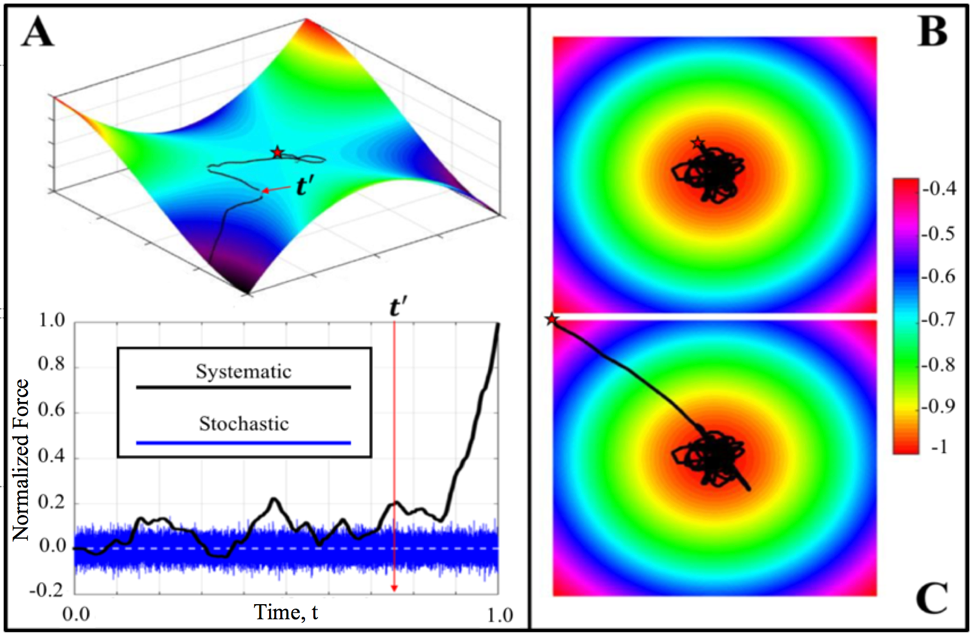}
\end{center}

\noindent 

\noindent \textbf{\textit{Figure 1. Schematic plots of typical Langevin trajectories}}\textit{. For each scenario }\textbf{\textit{(A-C)}}\textit{, the red star indicates the starting location. }\textbf{\textit{(A)}}\textit{ A particle's trajectory is initialized at a saddle point, and shown to escape for a short time. Below, the normalized systematic and stochastic forces are plotted as a function of time, t. The systematic force is the dominating effect when the particle locates the stronger potential gradient, as indicated at time }$t=t^{\mathrm{'}}$\textit{. }\textbf{\textit{(B)}}\textit{ A particle's trajectory is initialized near a minimum, where it is contained. }\textbf{\textit{(C)}}\textit{ A particle's trajectory is initialized away from a minimum, but ends up contained within the minimum. }

\subsection{Asymptotic matching of quantum and classical distributions}

\noindent As others have noted \cite{ARTICLE:16, ARTICLE:17, ARTICLE:22, ARTICLE:24, ARTICLE:28}, the conventional quantum mechanical probability density function, ${\rho }_B\left(\mathbf x\right)={\left|{\psi }_0\left(\mathbf x\right)\right|}^2$, can be interpreted as an empirical probability distribution obtained from the original data points, ${\left\{{\mathbf x}_i\right\}}^N_{i=1}$. However, we assume that this data set is a sample drawn from a more general distribution that obeys hypotheses of a canonical ensemble. We emphasize that this is the most distinguishable assumption in this method: instead of viewing ${\left\{{\mathbf x}_i\right\}}^N_{i=1}$ as a closed system, we consider it as a subsystem taken from an environment. As detailed above, such an environment is characterized by a temperature, $T$, and there exists a critical temperature, $T_0$, such that the quantum distribution is asymptotically convergent to a classical distribution (with a discrepancy of order $\mathrm{O}\left(\varepsilon \right)$). As such, in this section we aim to asymptotically match the quantum distribution with a classical distribution, and in the process derive an analytical expression for critical temperature, $T_0$. 

\noindent \textbf{}

\noindent In general, the Langevin dynamics lead to an ergodic trajectory with respect to the classical Gibbs distribution,
\begin{equation}F\left(\mathbf q,\mathbf p\right)\propto e^{-\beta H\left(\mathbf q,\mathbf p\right)}\equiv e^{-\beta \left(\frac{{\left|\mathbf p\right|}^2}{2}+V\left(\mathbf q\right)\right)},\end{equation}

\noindent whose marginal distribution is given by,

\noindent \textbf{}
\begin{equation}f\left(\mathbf q\right)=\int{d\mathbf p\, F\left(\mathbf q,\mathbf p\right)\propto }e^{-\beta V\left(\mathbf q\right)}.\end{equation} 

\noindent \textbf{}

\noindent We then denote the partition function for this marginal distribution as $Z$, such that,

\noindent 
\begin{equation}f\left(\mathbf q\right)=\frac{1}{Z}e^{-\beta V\left(\mathbf q\right)},\ \ \ \int{f\left(\mathbf q\right)}d\mathbf q=1.\end{equation} 

\noindent We have thus derived the approximate probability density function from two different perspectives: quantum mechanical and classical mechanical. As such, we propose the following consistency condition, 
\begin{equation}{\rho }_A\left(\mathbf x\right)={\left|{\psi }_0\left(\mathbf x\right)\right|}^2\approx \frac{1}{Z}e^{-\beta V\left(\mathbf x\right)}.\ \ \ \end{equation} 

\noindent By definition, ${\psi }_0\ge 0$, such that Equation (29) is equivalent to,  

\noindent \textbf{} 
\begin{equation}{\psi }_0\left(\mathbf x\right)\approx \frac{1}{\sqrt{Z}}e^{-\frac{\beta }{2}V\left(\mathbf x\right)},\end{equation} 

\noindent \textbf{}
\noindent which leads to the following matching condition, 

\noindent \textbf{}
\begin{equation}C_N\sum^N_{i=1}{e^{-\frac{{\left|\mathbf x-{\mathbf x}_i\right|}^2}{2\varepsilon }}}\approx \frac{1}{\sqrt{Z}}e^{-\frac{\beta }{2}C}e^{-\frac{\beta }{2}\sum^N_{i=1}{w_i\left(\mathbf x\right)\frac{{\left|\mathbf x-{\mathbf x}_i\right|}^2}{2}}}.\end{equation}

\noindent This matching condition is carried out under the assumption that, given ${\left\{{\mathbf x}_i\right\}}^N_{i=1}$, one can choose a sufficiently small $\varepsilon $, such that the following non-overlapping wave-packet condition, 

\noindent 
\begin{equation}w_i\left({\mathbf x}_j\right)\approx {\delta }_{ij},\end{equation}

\noindent is nearly satisfied according to the weight functions defined via Equation (12). This non-overlapping wave-packet condition means that the set of wave-packets, ${\left\{{\psi }_i\right\}}^N_{i=1}$, composing the wavefunction, ${\psi }_0\left(\mathbf x\right)$, are nearly-isolated Dirac delta functions at time, $t=0$. 

\noindent \textbf{}

\noindent We then proceed to reduce the matching condition into two separate components: (1) matching function terms and (2) matching constant terms. First, if the function terms are to be matched at $\left\{{\mathbf x}_j\right\}$, the non-overlapping condition implies that for $\varepsilon \ll 1$, then $\beta =\frac{2}{\varepsilon }$. This necessary condition suggests that the absolute critical temperature is,

\noindent 
\begin{equation}T_0=\frac{1}{2}\varepsilon.\end{equation} 

\noindent We show in the following that when $\beta =\frac{2}{\varepsilon }$, not only are the two functions matched at $\left\{{\mathbf x}_j\right\}$, but they are approximately identity on the whole space. If we denote an effective support of the i${}^{th}$ wavepacket, ${\psi }_i$, as ${\mathrm{\Omega }}^{\varepsilon }_i$, we conclude that $w_i\left(\mathbf x\right)\approx 1$ when $\mathbf x\in {\mathrm{\Omega }}^{\varepsilon }_i$, otherwise $w_i\left(\mathbf x\right)\approx 0$. By using the identity $\sum^N_{i=1}{w_i\left(\mathbf x\right)}=1$, we have when $\beta =\frac{2}{\varepsilon }$,

\noindent \textbf{}
\begin{equation}e^{-\frac{\beta }{2}\sum^N_{i=1}{w_i\left(\mathbf x\right)\frac{{\left|\mathbf x-{\mathbf x}_i\right|}^2}{2}}}=\sum^N_{j=1}{{w_j\left(\mathbf x\right)e}^{-\sum^N_{i=1}{w_i\left(\mathbf x\right)\frac{{\left|\mathbf x-{\mathbf x}_i\right|}^2}{2\varepsilon }}}}\approx \sum^N_{j=1}{e^{-\frac{{\left|\mathbf x-{\mathbf x}_j\right|}^2}{2\varepsilon }}}.\end{equation} 

\noindent \textbf{}

\noindent Thus, we have shown that the function component of the matching condition is satisfied.

\noindent \textbf{}

\noindent Next, to match the constant terms, we note that since $C_N$ and $Z$ are both normalization constants, 

\noindent \textbf{} 
\begin{equation}{\left(C_N\right)}^2={\int{\left(\sum^N_{i=1}{e^{-\frac{{\left|\mathbf x-{\mathbf x}_i\right|}^2}{2\varepsilon }}}\right)}}^2d\mathbf x\approx \int{\left(e^{-\sum^N_{i=1}{w_i\left(\mathbf x\right)\frac{{\left|\mathbf x-{\mathbf x}_i\right|}^2}{\varepsilon }}}\right)}d\mathbf x=Ze^{\beta C}.\end{equation} 

\noindent \textbf{}
\noindent Further, the non-overlapping condition implies that when $i\neq j$, the wave-packets are approximately orthogonal, i.e.,

\noindent \textbf{}

\begin{equation}\int{\left(e^{-\frac{{\left|\mathbf x-{\mathbf x}_i\right|}^2}{2\varepsilon }}\right)\left(e^{-\frac{{\left|\mathbf x-{\mathbf x}_i\right|}^2}{2\varepsilon }}\right)}d\mathbf x\approx 0.\end{equation} 
 
\noindent \textbf{}

\noindent Then, by direct calculation, we get,

\begin{equation}{\int{\left(\sum^N_{i=1}{e^{-\frac{{\left|\mathbf x-{\mathbf x}_i\right|}^2}{2\varepsilon }}}\right)}}^2d\mathbf x\approx \int{\left(\sum^N_{i=1}{e^{-\frac{{\left|\mathbf x-{\mathbf x}_i\right|}^2}{\varepsilon }}}\right)}d\mathbf x={\left(\pi \varepsilon \right)}^k\sum^N_{i=1}{1}=N{\left(\pi \varepsilon \right)}^k,\end{equation} 

\noindent \textbf{}

\noindent which is the quantum mechanical normalization constant for a Gaussian wavefunction in k dimensions. Also, due to the non-overlapping condition, we know that ${\left\{{\mathbf x}_i\right\}}^N_{i=1}$ are approximate zeros of $V\left(\mathbf x\right)$. By the method of deepest decent,
\begin{equation}\int{e^{-\sum^N_{i=1}{w_i\left(\mathbf x\right)\frac{{\left|\mathbf x-{\mathbf x}_i\right|}^2}{\varepsilon }}}}d\mathbf x\approx {\left(\pi \varepsilon \right)}^k\sum^N_{i=1}{\left(1+\mathrm{O}\left(\varepsilon \right)\right)}.\end{equation} 
Therefore, we have verified that when $\beta =\frac{2}{\varepsilon }$, 
\begin{equation}{\rho }_A\left(\mathbf x\right)={\left|{\psi }_0\left(\mathbf x\right)\right|}^2\approx \frac{1}{Z}e^{-\beta V\left(\mathbf x\right)}={\rho }_B\left(\mathbf x\right),\end{equation}

\noindent and the quantum distribution has been approximately matched with the classical Gibbs distribution (i.e., Equations (1) and (7) are asymptotically matched at the critical temperature, $T=\frac{1}{2}\varepsilon $). 

\noindent \textbf{}

\noindent The illustrating example in \textbf{Figure 2} demonstrates quantum and classical probability distributions matched at critical temperature, $T_0={1}/{{\beta }_0}$ (Equation (33)), by using two 1D Gaussian wave-packets, ${\psi }_i$ and ${\psi }_j$, separated by various distances along the x-axis. When the non-overlapping condition (Equation (32)) is either completely or nearly satisfied based on negligible wave-packet interference, we demonstrate that the quantum and classical probability distributions are in reasonably good agreement. However, this agreement fails when the overlap between the two wave-packets becomes large, such that $w_i\left({\mathbf x}_j\right)\neq {\delta }_{ij}$. 

\begin{center}
%\noindent \includegraphics*[width=5.85in, height=3.50in, keepaspectratio=false]{image2}
\includegraphics[width=.9\textwidth]{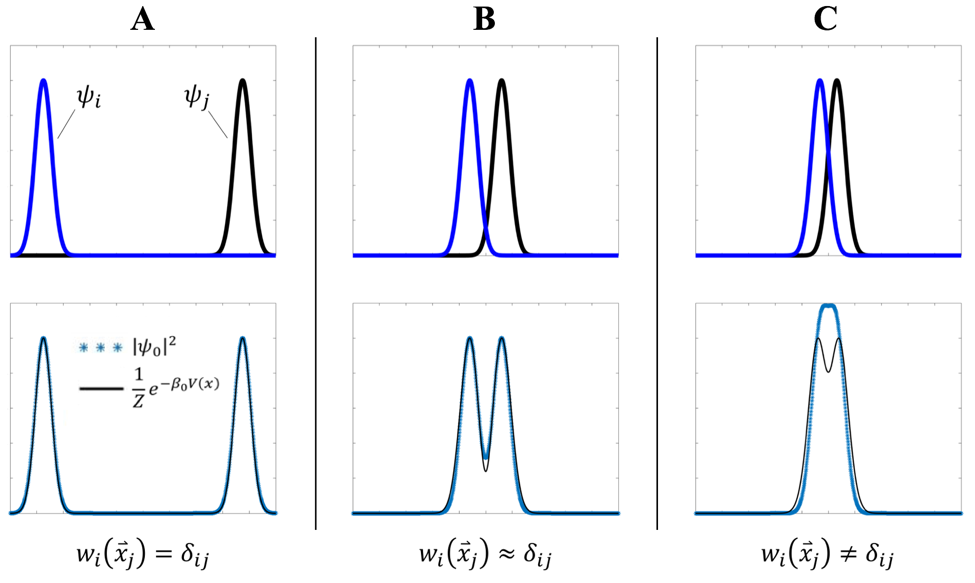}
\end{center}
	
\noindent \textbf{\textit{Figure 2.}}\textit{ }\textbf{\textit{Illustration of}}\textit{ }\textbf{\textit{quantum and classical probability distributions matched at critical temperature, }}${\boldsymbol{T}}_{\boldsymbol{o}}\boldsymbol{=}{\boldsymbol{\beta }}^{\boldsymbol{-}\boldsymbol{1}}_{\boldsymbol{o}}$\textbf{\textit{.}}\textit{ Two 1D Gaussian wave-packets, }${\psi }_i$\textit{ and }${\psi }_j$\textit{, are separated by various distances along the horizontal axis }\textbf{\textit{(top)}}\textit{.}\textbf{\textit{ }}\textit{The probability distributions }\textbf{\textit{(bottom) }}\textit{are in good agreement when the non-overlapping condition is either: }\textbf{\textit{(A)}}\textit{ completely satisfied, i.e., }$w_i\left({\mathbf x}_j\right)={\delta }_{ij}$\textit{, or }\textbf{\textit{(B)}}\textit{ nearly satisfied, i.e., }$w_i\left({\mathbf x}_j\right)\approx {\delta }_{ij}$\textit{. }\textbf{\textit{(C)}}\textit{ The probability distributions do not agree when the distance between the two wave-packets becomes small, where the non-overlapping condition is not satisfied, i.e., }$w_i\left({\mathbf x}_j\right)\neq {\delta }_{ij}$\textit{.}

\section{Time Integrator for Langevin Dynamics}

\noindent Let us recall for convenience the Langevin dynamics that are used to propagate the position and momentum of each wave-packet,

\noindent \textbf{}

\begin{equation}d\mathbf q=\mathbf pdt,\end{equation} 

\noindent and

\noindent \textbf{}

\begin{equation}d\mathbf p=-{\mathrm{\nabla }}_qV\left(\mathbf q\right)dt-\gamma \mathbf pdt+\sqrt{2{\gamma \beta }^{-1}}dB\end{equation}

\noindent where, $\mathbf q,\mathbf p\in {\mathbb{R}}^k$ are k-dimensional vectors of instantaneous position and momenta, respectively. To numerically integrate the Langevin dynamics, we apply the \textit{BAOAB} splitting scheme. This is a numerical integration technique previously developed by \cite{ARTICLE:34} to address classical molecular dynamics problems regarding a Langevin thermostat. Here, the Langevin dynamics are decomposed into three components: a kinetic component (denoted by ``A''), 

\noindent \textbf{}
\begin{equation}\left(d\mathbf q,d\mathbf p\right)=\left(\mathbf pdt,\ 0\right),\end{equation} 

\noindent \textbf{}

\noindent a potential component (denoted by ``B''),

\begin{equation}\left(d\mathbf q,d\mathbf p\right)=\left(0,\ -{\mathrm{\nabla }}_qV\left(\mathbf q\right)dt\right),\end{equation}

\noindent and a Langevin thermostat component (denoted by ``O'', for the Ornstein-Uhlenbeck equation),

\noindent \textbf{} 
\begin{equation}\left(d\mathbf q,d\mathbf p\right)=\left(0,-\gamma \mathbf pdt+\sqrt{2{\gamma \beta }^{-1}}dB\right).\end{equation} 

\noindent \textbf{}

\noindent The BAOAB procedure is then carried out as follows. The integration is discretized into \textbf{five} symmetric phases per discrete time-step, $\Delta t$: 

\noindent 

\begin{enumerate}
\item  the potential component (B) is integrated over $\frac{1}{2}\Delta t$ 

\item  the kinetic component (A) is integrated over $\frac{1}{2}\Delta t$

\item  the Langevin thermostat component (O) is integrated over $\Delta t$

\item  the kinetic component (A) is integrated over $\frac{1}{2}\Delta t$ 

\item  the potential component (B) is integrated over $\frac{1}{2}\Delta t$ 
\end{enumerate}

\noindent 

\noindent As each component is separable in position and momentum, each can be integrated explicitly, leading to high accuracy, reliable stability, and low computational cost \cite{ARTICLE:34, ARTICLE:35}. The kinetic and potential components correspond to drift and kick, respectively, as a result of global environmental interactions according to Equation (23) \cite{ARTICLE:34}. The Langevin thermostat component, defined via Equation (44), has the following Ornstein-Uhlenbeck solution,

\noindent \textbf{}
\begin{equation}\left(\mathbf q(t),\mathbf p(t)\right)=\left(\mathbf q\left(0\right),\ \ \ e^{-\gamma t}\mathbf p(0)+\sqrt{\left(1-e^{-2\gamma t}\right)\left({\beta }^{-1}\right)}\mathcal{N}\right)\end{equation}

\noindent 

\noindent where, $\mathcal{N}$\textbf{ }is a k-dimensional, uncorrelated Gaussian random field. This component governs local interactions, making it possible for local minima to merge as a consequence of thermal tunneling phenomena. As the individual components of the \textit{BAOAB} scheme essentially reduce to matrix level operations with analytic solutions, the entire algorithm is highly parallelizable, making it an ideal candidate for modern graphics-card enabled hardware. 

\section{Numerical Results and Validation}

\noindent In this section, we investigate the feasibility of data clustering with Langevin dynamics, using a standard benchmark dataset. Several computational experiments have been conducted to both validate the proposed methodology, and to better understand the choice of parameters driving clustering. \textit{Section 4.1}\textbf{ }introduces the benchmark dataset (and subsequent pre-processing steps) that was used to produce the subsequent numerical results, and \textit{Section 4.2} investigates clustering performance and hyper-parametrization. 

\subsection{Benchmark data and pre-processing}

\subsubsection{Ripley's crab data}

\noindent Ripley's Crab Data \cite{ARTICLE:36} was used as the basis of numerical experimentation and algorithm testing in this section. This is a standard dataset commonly used as a benchmark for classification problems. The 5-dimensional data contains 200 samples of \textit{Leptograpsus variegatus }crabs, each labeled as one of the following classes: (A) blue male, (B) blue female, (C) orange male, (D) orange female. Each sample consists of 5 morphological features -- one for each dimension of the problem -- which collectively capture sample-specific, quantitative information. 

\noindent \textbf{}

\noindent Accordingly, each sample is mathematically represented by a 5-dimensional feature vector, \[
{\left\{{\mathbf x}_i\right\}}^{200}_{i=1}\subset {\mathbb{R}}^5.\]
 The data is readily available in MATLAB (Mathworks, Natick, MA), where it is stored as a 200x5 matrix, $X$. Columns of $X$ correspond to a particular feature, ${\mathbf x}_j$, obtained from all samples, and rows of $X$ correspond to the 5-dimensional feature vectors, ${\mathbf x}_i$, representing each sample. To ensure normalized units, each column vector, ${\mathbf x}_j$, was zero-mean centered. This is a common pre-processing step in many statistical learning approaches, in particular, Principle Component Analysis (PCA) \cite{ARTICLE:37}. In general, the goal of the benchmark dataset is to reconstruct the categorical classes, based exclusively on the quantitative information encoded within the feature vectors.

\subsubsection{Singular value decomposition}

\noindent Rather than clustering $X$ in its original 5-dimensional basis, we first perform a singular value decomposition (SVD),
\begin{equation}X=\sum^5_{j=1}{{{\sigma }_j\mathop{u}\limits^{\rightharpoonup}}_j{{\mathop{v}\limits^{\rightharpoonup}}_j}^{\dagger }}\end{equation}

\noindent where, ${\mathop{u}\limits^{\rightharpoonup}}_j$ and ${\mathop{v}\limits^{\rightharpoonup}}_j$ are the left-and-right singular vectors of $X$, and ${\sigma }_j$ are the singular values of $X$. The vectors, ${\mathop{u}\limits^{\rightharpoonup}}_j$, define a practical 5-dimensional coordinate system for the data that is particularly useful for clustering problems \cite{ARTICLE:24, ARTICLE:28, ARTICLE:38}. To reduce the number of dimensions, we truncate Equation (46) to its first 3 leading terms, and cluster the points, ${\left\{{\mathop{y}\limits^{\rightharpoonup}}_i\right\}}^{200}_{i=1}\subset {\mathbb{R}}^3$, within the 3-dimensional space defined by the Principal coordinate axes, $({\mathop{u}\limits^{\rightharpoonup}}_1,{\mathop{u}\limits^{\rightharpoonup}}_2,{\mathop{u}\limits^{\rightharpoonup}}_3)$. 

\noindent \textbf{}

\noindent In general, hyper-dimensional statistical learning algorithms are often subject to overfitting artifacts as a result of too many degrees-of-freedom built into models. In fact, feature selection is a common and important first step to any modern machine learning approach. It is highly likely that our proposed algorithm should also benefit from reducing the number of dimensions in a given dataset. In particular, as the number of dimensions grows, the ability to find optimal values for $\varepsilon$ and $T$ becomes increasingly non-trivial\footnote{ In general, the optimization of $\varepsilon$  and   $T$ in arbitrary hyper-dimensions is a challenging task, and is a key focus of our future work.  }. The authors emphasize, however, that working in the truncated SVD space is only a pre-processing procedure used to reduce the dimensionality of the data and select a stable basis. While it provides a convenient coordinate basis to work in, SVD is not inherent to the clustering process itself. Other, possibly non-linear, basis transformations may also provide fruitful results. Further, as others have noted \cite{ARTICLE:16, ARTICLE:24}, a well-defined separation is observed in Ripley's Crab Data between the ${2}^{nd}$ and ${3}^{rd}$ Principal dimensions of the SVD space, i.e., ${(\mathop{u}\limits^{\rightharpoonup}}_2,{\mathop{u}\limits^{\rightharpoonup}}_3)$. In order to provide a comparison with the literature, many of the figures that follow are displayed across these Principal axes.

\subsection{Clustering performance}

\noindent In this section, we use Ripley's Crab Data to demonstrate data clustering via Langevin dynamics at sub-critical temperature. The authors stress that the data has been \textit{retrospectively} color-coded to differentiate known classes. This was done for demonstration purposes in order to easily monitor the dynamic clustering process. The algorithm, however, was blind to the data classes.

\noindent 

\noindent Different potential functions at various scales were constructed according to Equation (16) with semi-classical parameters ranging from $\varepsilon =0.0001$ to $\varepsilon =0.01$. For each resolution scenario, the data points, ${\left\{{\mathop{y}\limits^{\rightharpoonup}}_i\right\}}^{200}_{i=1}\subset {\mathbb{R}}^3$, were propagated according to Equations (21) and (22) at sub-critical temperature, $T={0.01T}_0=(0.01)(0.5\varepsilon )$. Each $\varepsilon $-specific dynamic process was run for a total time, $t_{max}\mathrm{=}10$, and a time step, $dt\mathrm{=}0.1$.

\noindent \textbf{}

\noindent Clustering performance was evaluated based on a generalized Jaccard score, $J$. This is a similarity metric comparing the algorithm-produced clustering results with known class labels obtained \textit{a priori} from the benchmark metadata \cite{ARTICLE:39}. Let $\boldsymbol{a}=\left(a_1,a_2,\dots ,a_{200}\right)$ be a vector whose elements correspond to the known class labels, and let $\boldsymbol{b}=\left(b_1,b_2,\dots ,b_{200}\right)$ be a vector whose elements correspond the clustering results, such that $a_i,b_i\in \ \left\{1,2,3,4\right\}\subset \mathbb{Z}$. The Jaccard score is then,
\begin{equation}J\left(\boldsymbol{a},\boldsymbol{b}\right)=\frac{\sum_i{min\left(a_i,b_i\right)}}{\sum_i{max\left(a_i,b_i\right)}}.\end{equation}

\noindent Essentially, J is a generalization of the receiver operator characteristic, extended to non-binary classification problems. Here, it is used to quantify the number of data points appearing in the same class according to both the algorithm and ground truth. A Jaccard score of $J=1$ indicates perfect clustering.  

\noindent \textbf{}

\noindent \textbf{Figure 3A} shows the Jaccard Score achieved for the range of $\varepsilon $ values at the sub-critical temperature, $T={0.01T}_0$, for a total time, $t_{max}\mathrm{=}10$, and a time step, $dt\mathrm{=}0.1$. As indicated by the red dashed vertical line, the performance of the algorithm was maximized at $\varepsilon =1.225E-3$, resulting in $J=0.90$. \textbf{Figure 3B} demonstrates a projection of the potential (the grayscale contour map) at this resolution, as well as the final location of each data point according to Langevin dynamics. Reasonable clustering performance ($J\ge 0.6$) was archived for a relatively stable range of $\varepsilon $ values within at the semi-classical scale, as indicated by the curve's broad shoulder between 0.001 and 0.005. As $\varepsilon \to 0$, performance is reduced due to decreasing-interaction among data points, and is demonstrated as $J$ drops to 0.3 at very low $\varepsilon $. Similarly, as $\varepsilon \to 1$, quality clustering is no longer achieved. In this non-semi-classical limit, the Jaccard score tappers to a sub-optimal $J=0.3$. Further, we note that clustering at critical temperature only produced a maximum Jaccard score of $J=0.72$, and that temperatures of $T\le 0.01T_0$ resulted in particularly stable clustering results. 

\noindent 

\begin{center}
	\includegraphics[width=.75\textwidth]{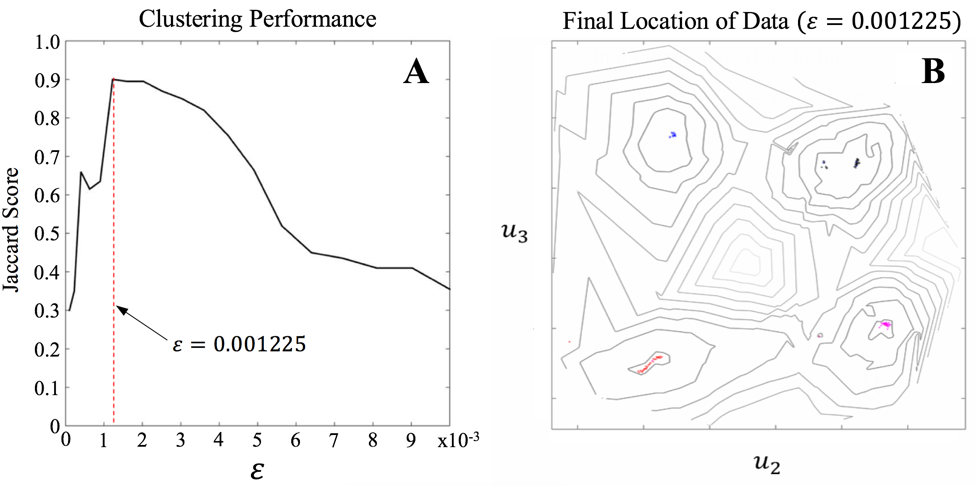}
\end{center}

\noindent \textbf{\textit{Figure 3.}}\textit{ }\textbf{\textit{Sensitivity of}}\textit{ }\textbf{\textit{clustering performance to different }}$\boldsymbol{\varepsilon }$\textbf{\textit{ values.}}\textit{ }\textbf{\textit{(A)}}\textit{ Jaccard scores, indicating clustering performance, were calculated at a sub-critical temperature of }$T={0.01T}_0$\textit{. Clustering was optimized at }$\varepsilon =0.001225$\textit{, resulting in a maximum Jaccard score of }$J=0.90$\textit{. }\textbf{\textit{(B)}}\textit{ The final location of each data point following Langevin dynamics (plotted in color), relative to a projection of the potential surface (gray-scale contour plot) at }$\varepsilon =0.001225$\textit{. Each color corresponds to a different data class, demonstrating optimal separation at distinct potential minima. Here, }$u_2$\textit{ and }$u_3$\textit{ represent the 2${}^{nd}$ and 3${}^{rd}$ Principal Component axes, respectively.}

\noindent \textbf{}

\noindent \textbf{}

\noindent The following figures, (\textbf{Figure 4}, \textbf{Figure 5} and \textbf{Figure 6}), all demonstrate different aspects of the dynamic clustering process with input parameters: $(\varepsilon ,T)=(1.225E-3,\ \ 0.01T_0)$. Specifically, \textbf{Figure 4} shows the time-evolution of the data as a series of sequential snap-shots, comparing the dynamics between critical (\textbf{Figure 4A}) and sub-critical temperatures (\textbf{Figure 4B}). The potential function is shown as a 2D projection across the 2${}^{nd}$ and 3${}^{rd}$ Principal Component axes, and the data points have been retrospectively color-coded according to the known data classes. At each sequential time-evolution frame, the difference between critical and sub-critical evolution becomes more-and-more apparent. The sub-critical Langevin dynamics were run at a resolution of $\mathrm{\varepsilonup }\mathrm{=}1.225E-3$, with a temperature of $\mathrm{T=0.01}{\mathrm{T}}_0$, a total time, $t_{max}\mathrm{=}10$, and a time step, $dt\mathrm{=}0.1$. While data aggregation is easily observed at $T=0.01T_0$, the Langevin dynamics at critical temperature, $T=T_0$, unveils a much less obvious structure to the dataset. This further fulfills our intuition that clustering is a sub-critical temperature phenomenon. 

\noindent 

\begin{center}
	\includegraphics[width=.9\textwidth]{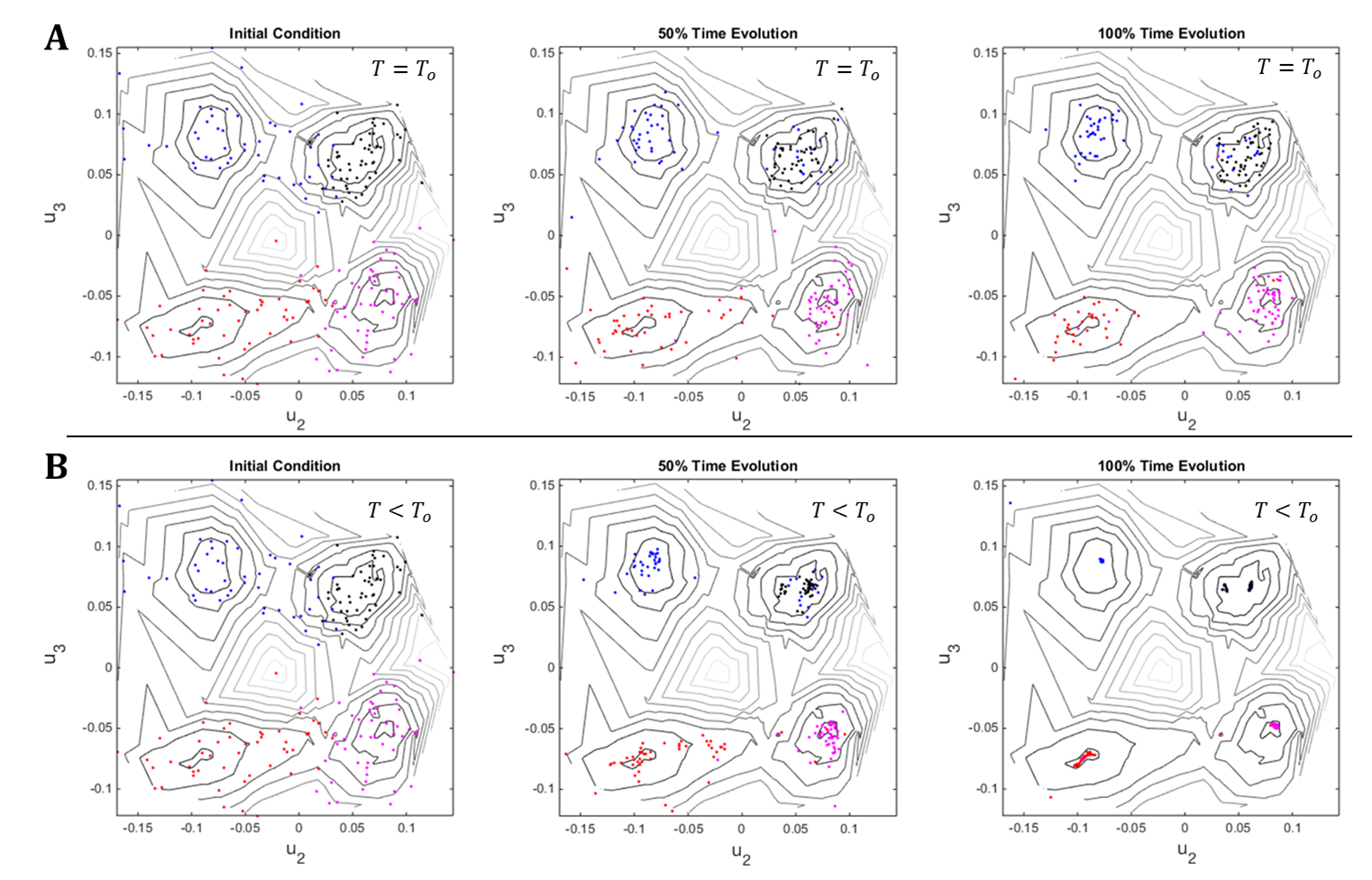}
\end{center}

\noindent \textbf{\textit{Figure 4.}}\textit{ }\textbf{\textit{Data clustering as a result of Langevin annealing}}\textit{. The time-dependent location of each data point is plotted, from left-to-right, relative to a projection of the potential surface at }$\varepsilon =0.001225$\textit{. Here, }$u_2$\textit{ and }$u_3$\textit{ represent the 2${}^{nd}$ and 3${}^{rd}$ Principal Component axes, respectively. }\textbf{\textit{(A) }}\textit{At critical temperature,}\textbf{\textit{ }}$T=T_0$\textit{,}\textbf{\textit{ }}\textit{the Langevin dynamics produce}\textbf{\textit{ }}\textit{a}\textbf{\textit{ }}\textit{Jaccard score of }$J=0.72$\textit{. }\textbf{\textit{(B) }}\textit{At a lower sub-critical temperature,}\textbf{\textit{ }}$T{<T}_0$\textit{,}\textbf{\textit{ }}\textit{the Langevin dynamics produce}\textbf{\textit{ }}\textit{a}\textbf{\textit{ }}\textit{Jaccard score of }$J=0.90$\textit{. Here, a clear separation of the data points is demonstrated, as each color corresponds to a different data class. }

\noindent \textbf{}

\noindent \textbf{}

\noindent Next, \textbf{Figure 5} shows the same time-evolution process as \textbf{Figure 4}, but follows the location of each data point across all 3 Principal coordinates of the truncated SVD space. Even though the known separation in the data classes is optimal across the 2${}^{nd}$ and 3${}^{rd}$ dimensions, \textbf{Figure 5} demonstrates that the spatial information encoded within the other dimensions is not lost during the clustering process. The location of the data points, which are again retrospectively color-coded according to class, are shown to converge towards similar coordinates. 

\noindent 

\noindent 

\begin{center}
	\includegraphics[width=1\textwidth]{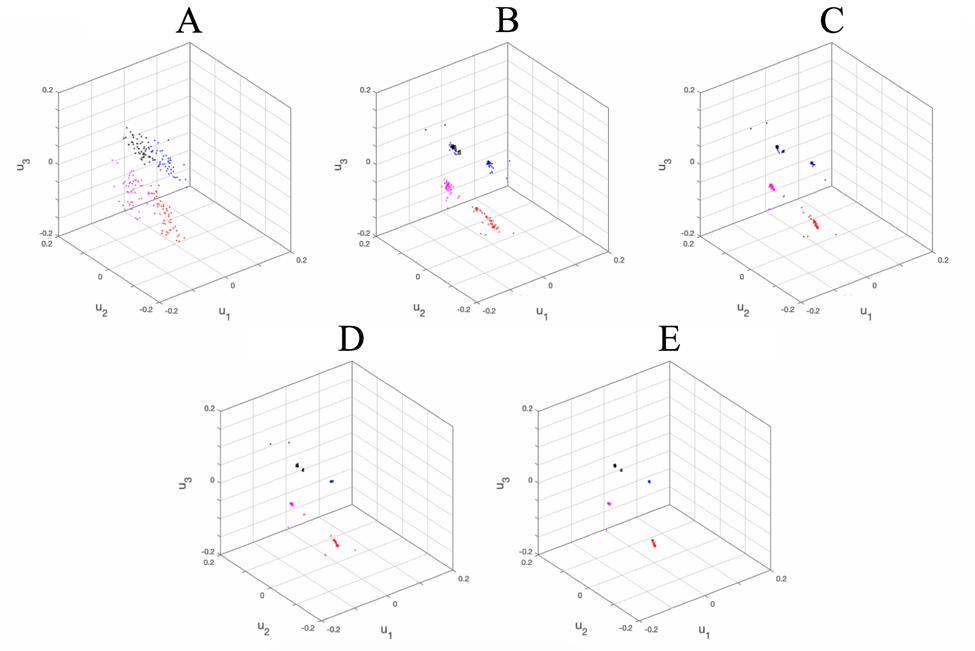}
\end{center}

\noindent 

\noindent \textbf{\textit{Figure 5.}}\textit{ }\textbf{\textit{Time-dependent location of data points}}\textit{ }\textbf{\textit{in 3 Principal dimensions.}}\textit{ The time-dependent location of each data point is plotted as a series of 5 sequential snap shots }\textbf{\textit{(A-E)}}\textit{ for }$\varepsilon =0.001225$\textit{ and }$T=0.01T_0$\textit{. Here, }$u_1$\textit{, }$u_2$\textit{ and }$u_3$\textit{ represent the 1${}^{st}$, 2${}^{nd}$ and 3${}^{rd}$ Principal Component axes, respectively. }\textbf{\textit{(E)}}\textit{ Data points of the same class (differentiated by color) are shown to aggregate at similar spatial locations as a result of the sub-critical Langevin dynamics, producing a Jaccard score of }$J=0.90$\textit{.  }

\noindent \textbf{}

\noindent \textbf{} 

\noindent Finally, \textbf{Figure 6} demonstrates the time-dependent changes of the weight matrix, $w_{ij}$, defined via Equation (25) , as a result of the Langevin dynamics. Clearly, at 10\% of the time-evolution, $w_{ij}$ is still densely populated along the main diagonal (thus satisfying Equation (32), the non-overlapping condition). As the dynamic process continues, however, the four data classes are observed within $w_{ij}$, demonstrated by high intensity regions extending from the diagonal. From 70\%-100\% of the time-evolution, 4 sets of similar row-vectors are observed within $w_{ij}$, indicating that the corresponding data points have saturated to similar locations as a result of the dynamics.  

\begin{center}
	\includegraphics[width=.8\textwidth]{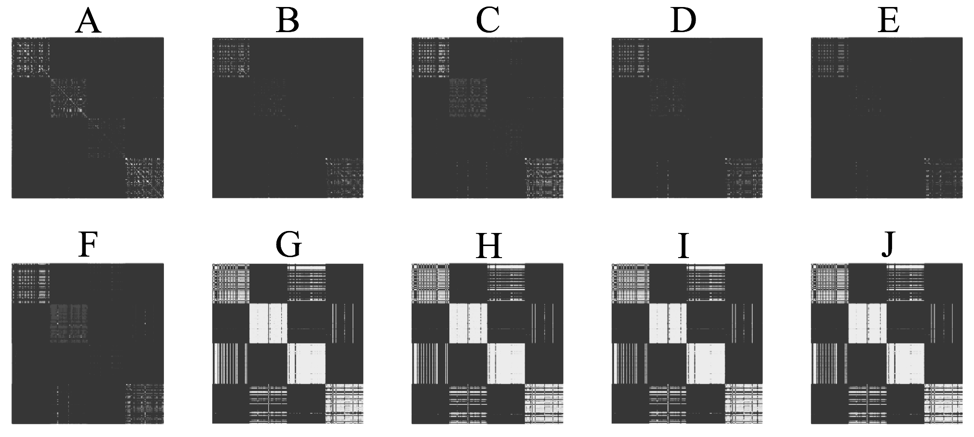}
\end{center}

\noindent \textbf{}

\noindent \textbf{\textit{Figure 6.}}\textit{ }\textbf{\textit{Data point overlap as a function of time evolution.}}\textit{ The Langevin dynamics for }$\varepsilon =0.001225$\textit{ and }$T=0.01T_0$\textit{ produce changes in data overlap, as measured by the weight matrix, }$w_{ij}$\textbf{\textit{ }}\textit{(Equation (25)). }\textbf{\textit{(A-J)}}\textit{ }$w_{ij}$\textit{ is shown as a series of 10 sequential snap shots in time, where the brightness (intensity) indicates the degree of data overlap. High intensity (white) matrix values of }$w_{ij}$\textit{ represent a significant overlap between the i${}^{th}$ and j${}^{th}$ data points. }\textbf{\textit{(G-J) }}\textit{The response of }$w_{ij}$\textbf{\textit{ }}\textit{to Langevin dynamics results in a saturated diagonal block behavior, capturing the 4 data classes. }

\section{Summary and Outlook}

\noindent In this paper, we have proposed a dynamic approach to data clustering based on fundamental concepts borrowed from both quantum mechanics and statistical mechanics. Due to the alarming rate at which \textit{big data} is being produced, the ability to understand hyper-dimensional datasets is becoming increasingly non-trivial. The development of novel clustering algorithms -- such as the one outlined in this paper -- is therefore an essential aspect of contemporary data analytics. 

\noindent \textbf{}

\noindent Our proposed methodology is briefly summarized as follows. A given dataset is interpreted as a classical ensemble in equilibrium with a heat bath. This produces an equilibrium distribution uniquely characterized by a temperature, $T$, and a resolution, $\varepsilon $. Langevin dynamics are then employed to stochastically propagate data points on ergodic trajectories obeying a classical Gibbs distribution according to the appropriate potential function. Data points which end up in the neighborhood of a particular potential minimum are ultimately considered to be in the same cluster. Time-evolution according to Langevin dynamics has several practical advantages with respect to robust data clustering. Most remarkably is the ability for data points to thermally jump local potential barriers and escape saddle points into locations of the potential surface otherwise forbidden. This tunneling phenomenon contributes to the overall dynamic optimization and ultimate partitioning of the data into degenerate subsets. 

\noindent \textbf{}

\noindent The clustering methodology was  applied to Ripley's Crab Data, a common 5-dimensional benchmark dataset often used in classification problems. Accurate clustering of the data classes was demonstrated at sub-critical temperatures. Such a critical temperature, $T_0$, was derived by asymptotically matching the quantum and classical distributions. Clustering performance was evaluated via the Jaccard Score, where the algorithm achieved a maximum value of 90\% at $T={0.01T}_0$ and $\varepsilon =1.225E-3$. At this same resolution, critical temperature Langevin dynamics produced a slightly worse Jaccard score of 72\%.  

\noindent \textbf{} 

\noindent Others have demonstrated that Langevin dynamics can be implemented to better optimize conventional statistical learning techniques, including Bayesian learning \cite{ARTICLE:42} and deep neural networks \cite{ARTICLE:43}. Our algorithm can potentially be improved in a similar manner by coupling it with traditional machine learning algorithms. In our case, the optimal scale to probe data is different for various datasets, and therefore choosing reliable $\varepsilon $ and $T$ values is fundamentally a task-based endeavor. Further, these parameters need to be chosen in a sequential way: $\varepsilon $ defines an \textit{ideal} potential landscape, and $T$ defines the dynamics that yield the \textit{desired} clustering results. Pairing our current approach with traditional statistical learning techniques would generalize such hyper-parameterization in a task-based manner. For example, implementation of a Bayesian feedback loop tertiary to the main algorithm would continuously improve the clustering performance as more data is warehoused to support a particular application. Thus, in principle, the methodology presented in this paper only represents the core component of a potentially much larger computational pipeline with a broad range of applications. 

\noindent \textbf{}

\noindent Other future research will consist of a detailed comparison study between our approach with other dynamic clustering algorithms. We also plan to (a) explore new challenges that arise when the number of dimensions grows significantly large; and (b) investigate the potential benefit of colored-noise Langevin dynamics, where a spatially-dependent temperature gradient may produce new, fruitful clustering results. 

\noindent \textbf{}

\noindent \textbf{\large Acknowledgments}

\begin{enumerate}
\item  K. Lafata is partially supported by a grant from Varian Medical Systems, Inc.

\item  Z. Zhou is partially supported by RNMS11-07444 (KI-Net) and the startup grant from Peking University.

\item  J. Liu is partially supported by KI-Net NSF RNMS grant No. 11-07444 and NSF grant DMS 1514826.

\item  F. Yin is partially supported by NIH 2R21CA218940, NIH 1R01CA184173, NIH 1R21CA165384, and a grant from Varian Medical Systems, Inc.
\end{enumerate}

\noindent 

\noindent \textbf{}

\bibliography{LangevinReferences} 
\bibliographystyle{ieeetr}

\end{document}